# Chemical heterogeneity enhances hydrogen resistance in high-strength steels

**Authors:** Binhan Sun[1*], Wenjun Lu[1,2], Ran Ding[3], Surendra Kumar Makineni[1,4], Baptiste Gault[1,5], Chun-Hung Wu[1], Di Wan[6], Hao Chen[3], Dirk Ponge[1] and Dierk Raabe[1*]

**Affiliations:**

[1]Max-Planck-Institut für Eisenforschung GmbH, Max-Planck-Straße 1, 40237 Düsseldorf, Germany.

[2]Department of Mechanical and Energy Engineering, Southern University of Science and Technology, 518055 Shenzhen, China.

[3]Key Laboratory for Advanced Materials of Ministry of Education, School of Materials Science and Engineering, Tsinghua University, 100084 Beijing, China.

[4]Department of Materials Engineering, Indian Institute of Science, 560012 Bangalore, India.

[5]Department of Materials, Royal School of Mines, Imperial College, Exhibition Road, SW7 2AZ London, UK.

[6]Department of Mechanical and Industrial Engineering, Norwegian University of Science and Technology, Richard Birkelands vei 2B, 7491 Trondheim, Norway.

*Correspondence to: b.sun@mpie.de; d.raabe@mpie.de.

**Abstract:** When H, the lightest, smallest and most abundant atom in the universe, makes its way into a high-strength alloy (>650 MPa), the material's load-bearing capacity is abruptly lost [1-3]. This phenomenon, known as H embrittlement, was responsible for the catastrophic and unpredictable failure of large engineering structures in service [4,5]. The inherent antagonism between high strength requirements and H embrittlement susceptibility strongly hinders the design of lightweight yet reliable structural components needed for carbon-free hydrogen-propelled industries and reduced-emission transportation solutions [1,6,7]. Inexpensive and scalable alloying and microstructural solutions that enable both, an intrinsically high resilience to H and high mechanical performance, must be found. Here we introduce a counterintuitive strategy to exploit typically undesired chemical heterogeneity within the material's microstructure that allows the local enhancement of crack resistance and local H trapping, thereby enhancing the resistance against H embrittlement. We deploy this approach to a lightweight, high-strength steel and produce a high-number density Mn-rich zones dispersed within the microstructure. These solute-rich buffer regions allow for local micro-tuning of the phase stability, arresting H-induced microcracks thus interrupting the H-assisted damage evolution chain, regardless of how and when H is introduced and also regardless of the underlying embrittling mechanisms. A superior H embrittlement resistance, increased by a factor of two compared to a reference material with a homogeneous solute distribution within each microstructure constituent, is achieved at no expense of the material's strength and ductility. Our strategy can be extended to other strong and ductile steels to increase their H-resistance at low cost. Exploiting chemical heterogeneities, rather than avoiding them, provides a new avenue for microstructure engineering via both, conventional thermomechanical processing and advanced metal processing techniques such as additive manufacturing.

**Main text:** Catastrophic failure induced by solute H is often caused by an accelerated damage evolution process involving (a) the ingress of typically only a few parts-per-million H, followed by its diffusion inside the microstructure interacting with various lattice defects (e.g. vacancies, dislocations and interfaces) [6,8], (b) the accelerated damage nucleation and propagation due to H-defect interactions [2,3,8-10] and (c) the accumulation of more H around the propagating crack tips through diffusion, driven by the crack-tip stress field [2,11], which further promotes cracking. Current engineering solutions normally include preventing H-ingress by applying protective coatings [3,12], which can fail under abrasive and corrosive environments [13]. An alternative is to design microstructures that are intrinsically resilient (e.g. through grain refinement [14] or introducing H-trapping precipitates [3,6,15]), yet, such measures have led to reduced strain-hardening ability and/or ductility in H-free condition [14,15]. Microalloying in steels to form various H-trapping precipitates (e.g. Ti- and V-based carbides) can suppress internal H migration [3,16], although it increases the materials' cost. However, this approach losses effectiveness when all H traps are filled to saturation [17,18], which can readily occur due to the typically low volume fraction of these precipitates (below 1% [19]), i.e. limited H storage volume. Steels represent ~90% of the global metallic alloy market, which means that even small improvements in their performance leverage worldwide impact. Yet, high-strength steels (e.g. used in the automotive industry for weight and carbon emission reduction) are extremely prone to H embrittlement, as less than 1 wt ppm H is often sufficient to result in a dramatic degradation of the mechanical properties [20-22]. It is thus particularly challenging to develop economical and scalable microstructure solutions to mitigate H embrittlement susceptibility of these materials, while maintaining their mechanical performance.

Here, we present a new strategy to tackle this challenge, through introducing solute heterogeneity in the steels' microstructure constituents. These well-designed local variations in composition enhance crack resistance locally, creating buffer zones which arrest H-induced microcracks that otherwise would rapidly propagate inside/along H-attacked phases or interfaces (Fig. 1a). We demonstrate our approach on a lightweight high-strength medium-Mn steel (0.2C-10Mn-3Al-1Si, in wt.%). A CALPHAD (Computer Coupling of Phase Diagrams and Thermochemistry)-assisted well-adjusted design of the Mn heterogeneity inside the austenite phase produces a high number density of microscopically confined Mn-rich buffer regions dispersed throughout the sample. During deformation of the alloy, the dynamic transformation from soft austenite to hard martensite is locally suppressed inside these buffer regions by the increased mechanical stability associated with their locally higher Mn content. As a result, the microstructure evolves into a high dispersion of softer islands embedded in the hard matrix, which renders H-induced microcracks frequently blunted and arrested. The chemical heterogeneity is here realized through a heat treatment with incomplete Mn partitioning/homogenization. The processing can be fully guided by the CALPHAD method and is readily scalable to established and affordable industrial processing routes. Also, it can be generalized to other types of high-strength steels with the same aim, namely, to lend their microstructures enhanced resistance against H embrittlement.

Our high-strength steel (strength level ~1 GPa) contains sub-micron sized domains of ferrite (α) and austenite (γ, Fig. 1b) and shows the transformation-induced plasticity (TRIP) effect, characterized by the deformation-driven displacive transformation from face-centered cubic austenite into body-centered cubic (or body-centered tetragonal) α'-martensite. The TRIP effect is a particularly efficient strain-hardening mechanism, often leading to the leap in mechanical properties [23] and driving the development of most modern advanced high-strength steels [23,24] and some maraging [25] and stainless steels [24,26]. However, its phase transformation product (martensite) and the associated hetero-interfaces are strongly prone to H-induced cracking, which is the major reason leading to premature failure of the entire material [11,15]. To manipulate the chemical heterogeneity inside the austenite phase, we designed a multi-step annealing process compatible with current industrial practice (see Methods). The process uses the slow kinetics of Mn homogenization between different Mn-containing austenite regions rapidly formed during a series of carefully designed annealing stages (Methods and Extended Data Fig. 1). A compositional variation of approx. 5 at.% Mn is produced inside austenite (Fig. 1c, d), which consists of Mn-rich stable buffer islands ($\gamma_{Mn-rich}$ containing 14~16 at.% Mn) and the surrounding Mn-lean metastable austenite regions ($\gamma_{Mn-lean}$ containing 11~12 at.% Mn). Multiple such Mn-rich zones are formed in single or multiple contiguous austenite grains (Fig. 1d). Taking advantage from the alternately arranged dual phase microstructure (Fig. 1b, c) and its ultrafine grain size (~0.6 μm), we achieved a high number density of $\gamma_{Mn-rich}$ (above $\sim 2\times 10^{18}$ m$^{-3}$) dispersed throughout the material. Its total volume fraction, however, remains below 5%, due to its small size (0.05~0.5 μm).

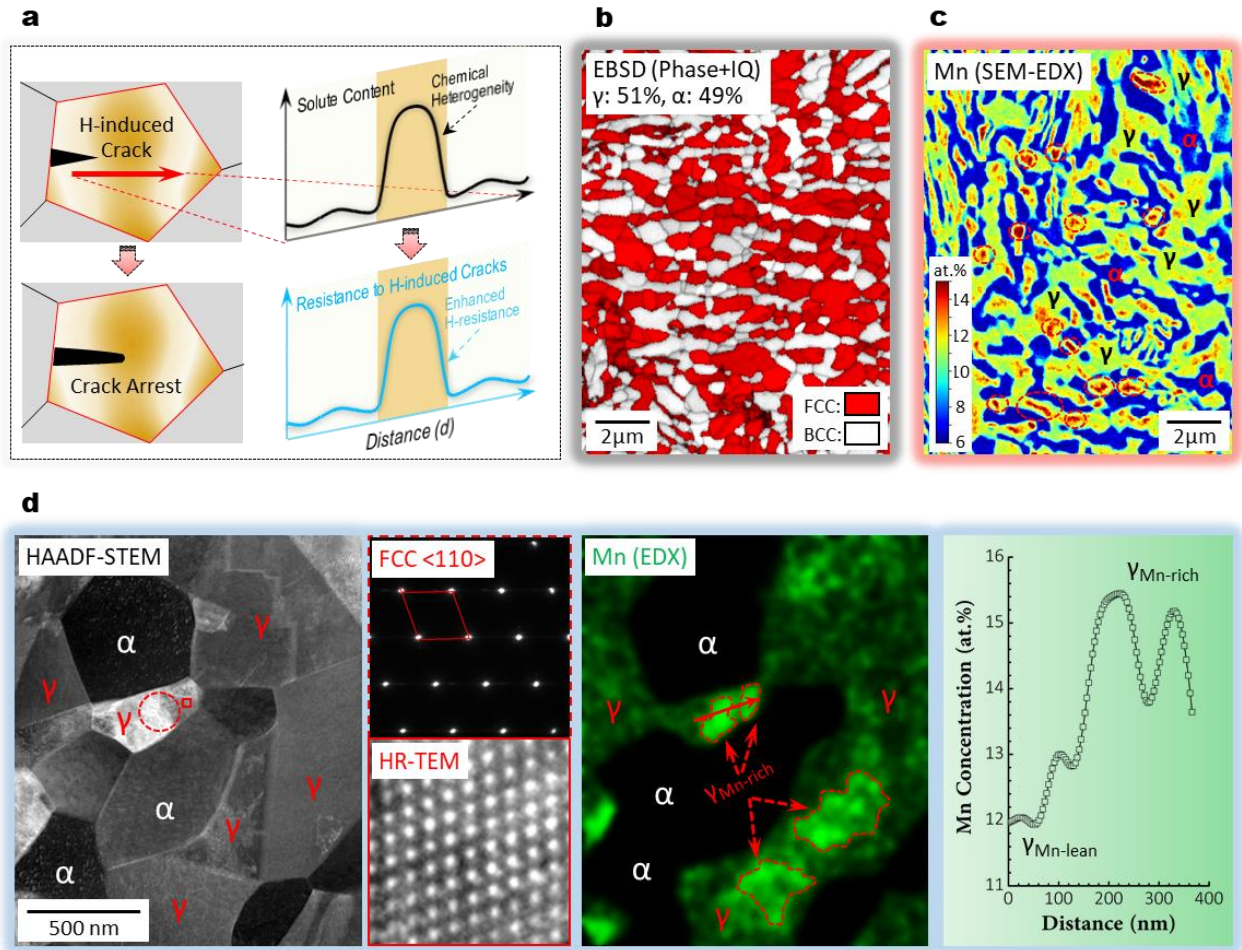

**Fig. 1 Concept of chemical heterogeneity-induced crack arresting as a measure against H embrittlement and the microstructure of a high-strength steel designed with a heterogeneous Mn distribution inside austenite. a,** Schematic image of the concept. **b,** Electron backscatter diffraction (EBSD) phase plus image quality (IQ) map showing the austenite-ferrite dual phase microstructure. **c,** Scanning electron microscopy (SEM) based energy-dispersive X-ray spectroscopy (EDX) map revealing the overall Mn distribution pattern in the microstructure. The chemical buffer zones are those regions where Mn is highly enriched (14~16 at.% Mn) inside the austenite phase (some of them are marked by elliptical frames). **d,** High-angle annular dark-field scanning transmission electron microscopy (HAADF-STEM) observation with EDX analysis, showing the existence of multiple Mn-rich zones inside one austenite crystal cluster or even one austenite grain. The selected area electron diffraction (SAED) and high-resolution TEM (HR-TEM) images taken respectively from the marked circular and rectangular frames are placed on the right side of the STEM image. The EDX line profile is taken from the area marked by an arrow in the EDX map.

Our approach of designing chemical heterogeneity results in a significantly improved H embrittlement resistance. This is revealed by comparing our steel having a heterogeneous Mn distribution (referred to as HET sample) to a reference specimen which has the same bulk

- 4 -

composition and a similar microstructure but processed to obtain a nearly homogeneous Mn distribution inside austenite (HOM sample, as shown in Extended Data Fig. 3 and Table 1). Both samples were subjected to the same H pre-charging condition, which yields a similar total H concentration (1~10 wt ppm controlled by varying the electrochemical charging conditions) and similar H penetration depth in the two samples, as suggested by thermal desorption spectroscopy (TDS, Extended Data Fig. 4) and permeation tests (see Methods), respectively. At the same H concentration, the HET sample shows nearly twice the total elongation (22.1~40.3%) compared with the HOM sample (10.3~22.3%), determined by slow strain rate (~$8\times10^{-5}$ s$^{-1}$) tensile testing (Fig. 2a). The mechanical performance under H-free condition, however, is nearly the same for the two samples (Fig. 2a and Extended Data Fig. 6). The strain hardening ability, controlled by the overall kinetics of deformation-induced austenite-to-martensite transformation (TRIP effect), is not affected by the presence of the small fraction of $\gamma_{Mn-rich}$ (below 5 vol.%), as demonstrated in Extended Data Fig. 6. This reveals the ability of the manipulated chemical heterogeneity to reconcile the most important, yet, often mutually exclusive material's features in this field, namely, high H-resistance and high strain-hardening. The latter is particularly important for maintaining ductility (and formability) in high-strength alloys, as it counteracts Considère-type plastic instability.

We further compare our microstructure engineering approach with other previously reported methods for enhancing the resistance to H embrittlement (Fig. 2b and Extended Data Fig. 5). The H-resistance is here characterized by both the tensile ductility in the presence of H and by the so-called H embrittlement index (the ratio of ductility among the samples tested with and without H [27]). While the introduction of H-trapping precipitates can improve the H-resistance before H-saturation, the improvement is typically below ~75% (Fig. 2b and Extended Data Fig. 5) which is bounded by the limited H storage volume. The formation of precipitates, even though confined to the nanoscale, can also result in detrimental effects, for instance, due to the introduced elasto-plastic deformation fields inside the adjacent matrix that increase the uptake of diffusible H [3,18]. Grain refinement improves the H embrittlement index generally in austenitic materials (Extended Data Fig. 5). However, the ductility in fine grained materials (especially with a grain size below 2 μm) can be seriously degraded in both H-containing and H-free conditions (Fig. 2b). This is due to the limited strain-hardening of such materials, which can promote early plastic instability associated with H-induced local plasticity (i.e. softening) [28]. Our approach does not suffer from these limitations and achieves a higher improvement in H embrittlement resistance (up to 122%), surpassing the effects achieved by the other two widely adopted strategies.

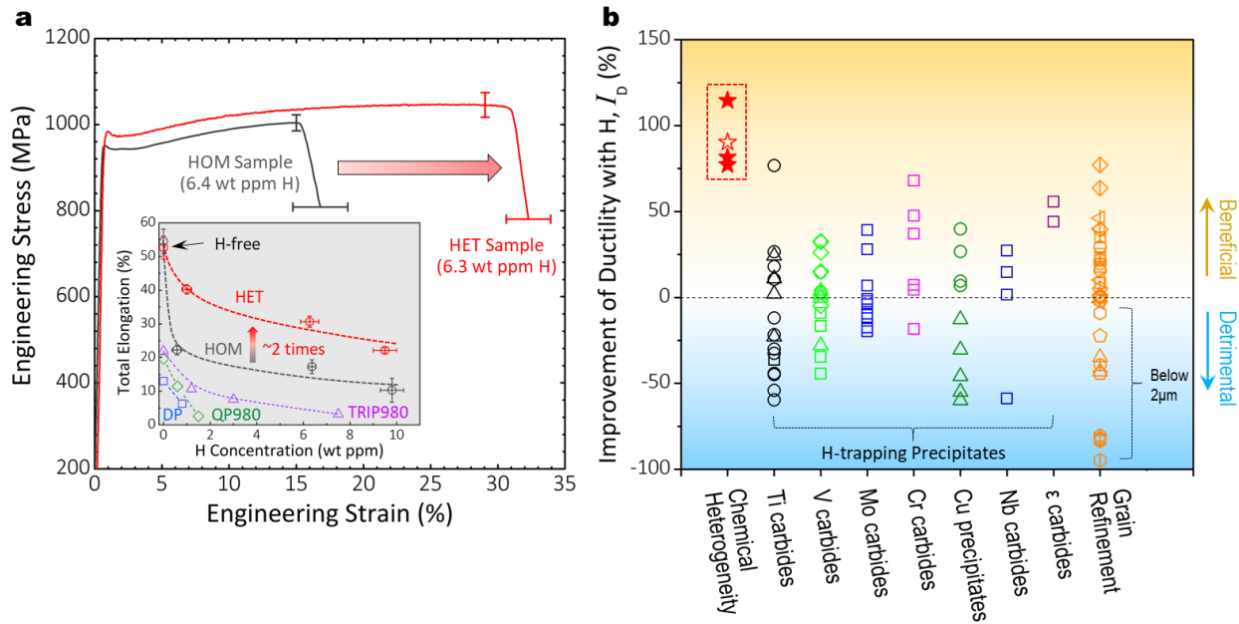

**Fig. 2 Improvement of the resistance to H embrittlement via designing chemical heterogeneity. a,** Slow strain rate tensile properties of the sample with a heterogeneous Mn distribution inside austenite (HET sample) in comparison to the reference material with homogenous solute distribution (HOM sample). Both samples were subjected to the same H pre-charging conditions (similar H amount). Representative tensile curves of samples with a H concentration of ~6.5 wt ppm are shown here. The inset shows the ductility of the two samples as a function of H concentration. The literature data tested on commercial high-strength steels (TRIP 980 [29], Q&P 980 [30] and DP steel [21]) with a similar strength level (i.e. yield strength above ~600 MPa and tensile strength of ~1000 MPa) are also included, serving as an indication for the required resistance to H embrittlement in automotive application. **b,** Comparison between our approach and other H-resistance enhancing methods reported in the literature [14,15,18,22,28,31-49], in terms of their effects on the improvement of the tensile ductility in the presence of H ($I_D=(El_{t-H,after}-El_{t-H,before})/El_{t-H,before}\times100\%$, where $El_{t-H,before}$ and $El_{t-H,after}$ are the total elongations of the H-containing specimens before and after applying the resistance-enhancing approach, respectively). Our tests performed on both, H pre-charged specimens (solid symbols) and specimens under continuous electrochemical charging during deformation (hollow symbol, results also shown in Extended Data Fig. 11), are shown. The literature data are from different materials including pure iron [36], low-alloy martensitic steels [18,32,33,35], pipeline steels [34,38], various advanced-high strength steels [14,15,22,31,37,45,47,49], maraging steels [48], various austenitic stainless steels [28,39,41,44,46] and high entropy alloys [42,43], subjected to electrochemical or gaseous $H_2$ pre-charging or *in-situ* H charging. The different symbols represent the data from different reports.

Next we validate the local resistance provided by the Mn-rich buffer zones ($\gamma_{Mn-rich}$) to H-induced microcracks and unveil the underlying crack-arresting mechanisms. First we focus on the deformation behavior of the austenite within the Mn-rich regions (Fig. 3). The higher Mn content increases the austenite's mechanical stability locally by reducing the chemical driving force for the austenite-to-martensite phase transformation. Therefore, unlike the austenite in other regions where

- 6 -

α'-martensite forms upon straining through the TRIP effect, the Mn-rich austenite zones resist phase transformation (Fig. 3a, b) and instead, deform through the gliding of partial dislocations and formation of nano-twins (see HR-TEM and SAED results in Fig. 3b) due to the increased stacking fault energy [50]. The capacity of these $\gamma_{Mn-rich}$ domains to maintain their phase stability turns them into plastically compliant buffer zones which can arrest H-induced cracks that intrude from the neighboring transformed regions. The arresting mechanism, as illustrated schematically in Fig. 4a, works through two effects: (a) austenite has higher H solubility and lower H diffusivity than α'-martensite (differing both by more than two orders of magnitude [20,51]), and it thus serves as a H-trapping zone slowing down H migration [3,11,20]; (b) the enhanced plastic compliance and flow of austenite at the crack tips results in crack blunting. The blunt cracks are frequently observed in the current H-charged and fractured HET sample and are often surrounded by untransformed austenite (Extended Data Fig. 8). In these austenite regions, EBSD analysis shows an increase in point-to-point misorientation angle close to the H-induced cracks, suggesting a strong plastic deformation at the propagating crack tips [52]. Such a crack blunting mechanism marks a distinct advantage of the current approach, as it operates even when all the traps are saturated by H, for instance after prolonged exposure to a H containing environment.

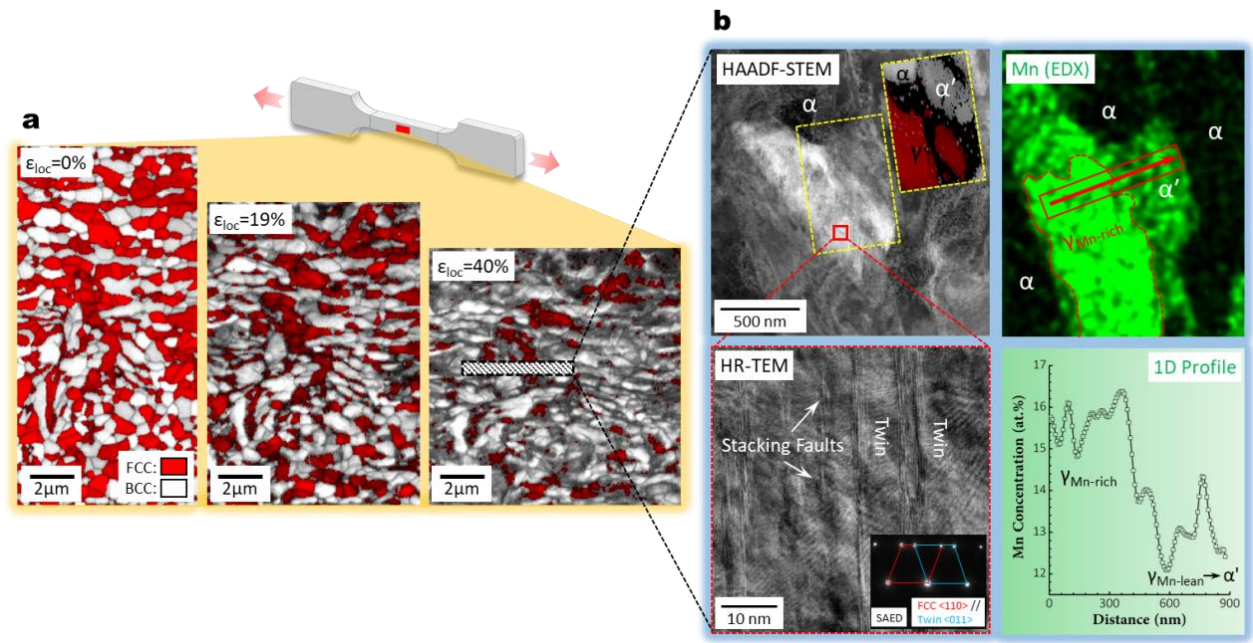

**Fig. 3 Microstructure evolution of the chemical heterogeneity-manipulated steel (HET sample) under H-free condition. a,** *Ex-situ* EBSD observation up to a local tensile strain ($\varepsilon_{loc}$) of 40%, showing the partial austenite-to-martensite transformation. **b,** STEM and EDX analysis generated roughly from the sub-surface of the rectangular frame marked in **(a)**. The inset in the HAADF-STEM image is a transmission Kikuchi

diffraction (TKD) phase plus IQ map, taken from the yellow rectangular frame. The EDX line profile is taken from the area marked by an arrow in the EDX map.

We used atom probe tomography (APT) to analyze the nanoscale composition in the region adjacent to a typical arrested H-induced crack in our heterogeneity-manipulated steel (Fig. 4b, c). The H-induced crack preferably nucleates at the interface between ferrite and α'-martensite (transformed from $\gamma_{Mn-lean}$ containing ~11 at.% Mn). Detailed APT data analysis of such hetero-interface reveals a tendency for H segregation (see Methods and Extended Data Fig. 7), which supports the role of H in promoting interface decohesion [53]. The nucleated crack is, however, arrested within the Mn-rich zone (~14 at.% Mn) where austenite resists load-driven phase transformation (Fig. 4b, c). Further crack propagation is thus no longer possible and, upon further deformation, new cracks would nucleate in other Mn-lean regions, a process with a high energy barrier [54] (e.g. the secondary crack shown in Fig. 4b). In some circumstances, the Mn-rich zones remain as ligaments bridging the crack wake, further reducing the crack-tip stress intensity and continuously suppressing the crack propagation [54].

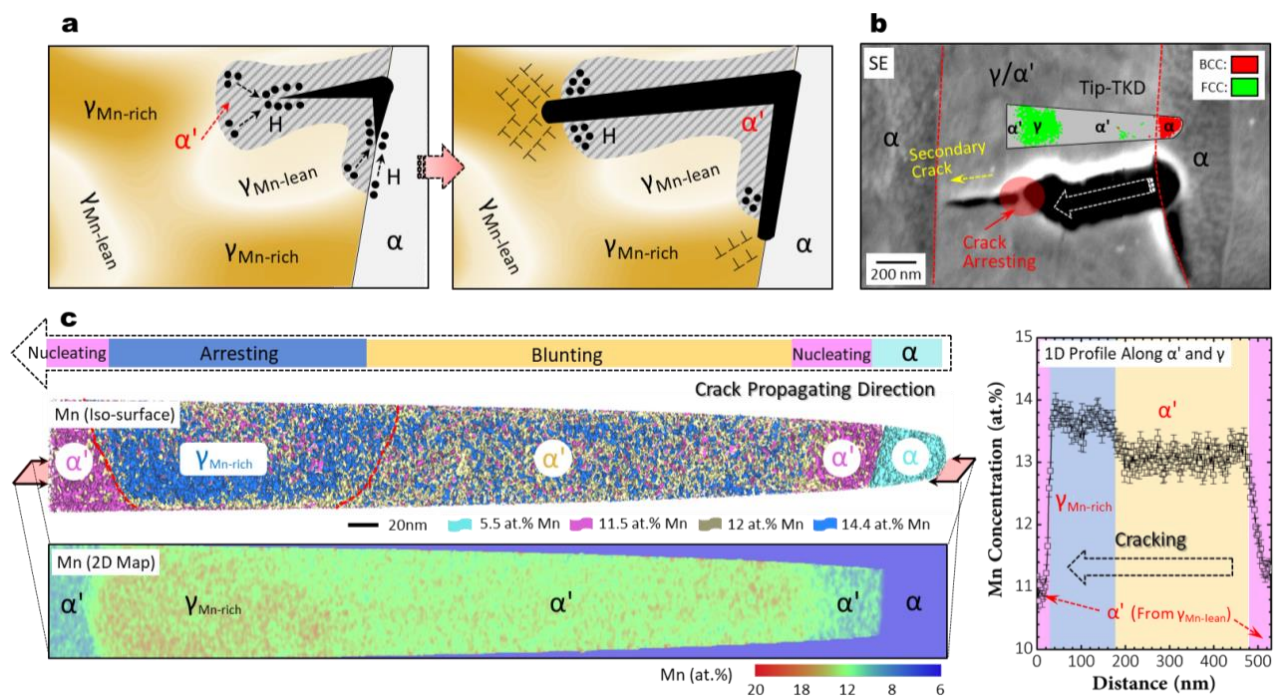

**Fig. 4 Chemical heterogeneity-induced arresting of H-induced cracks. a,** Schematic illustration showing the arresting mechanisms. **b,** A representative blunted and arrested H-induced crack in the H-charged and fractured chemical heterogeneity-manipulated steel (total H amount 6.3 wt ppm). The crack was imaged by secondary electron (SE) from the area close to the fracture surface and specimen side edge, where H is nearly saturated. The inset is the TKD phase map for the APT tip, which is placed at the region exactly where the

- 8 -

tip was lifted-out. **c,** The APT results for the tip shown in **(b)**, containing an iso-composition surface map for Mn, a two-dimensional Mn concentration map and a one-dimensional Mn profile across α' and γ.

The lattice structure change associated with the deformation-induced phase transformation brings the H atoms from their initial low-mobility solute state in the host austenite to a super-saturated and highly mobile state inside the product martensite [20,55,56]. This phase transformation-driven H release promotes damage evolution inside fresh martensite or along the associated hetero-interfaces [11]. In homogeneous TRIP-aided materials (e.g. the HOM sample), H-induced cracks propagate in an accelerated process, as the high stress/strain concentration near the nucleated crack tip triggers more α'-martensite formation (supported by Extended Data Fig. 9) thus liberating more mobile H. The crack arresting effect created by the dispersed Mn-rich buffer zones thus frequently interrupts this damage evolution chain, leading to superior H embrittlement resistance (Fig. 2). This is supported by the statistical crack analysis conducted on H-charged and fractured samples, which shows that the length of H-induced cracks near the fracture surface, regardless of their types (intergranular or transgranular), remains much shorter in the HET sample compared to the HOM counterpart (as quantified in Extended Data Fig. 10).

To further validate the enhancing effects of our approach on H-resistance, we also performed tests directly under H exposure, by *in-situ* electrochemical charging and *in-situ* H-plasma charging (see Methods). In both types of H containing environments, our approach demonstrates a greatly improved fracture resistance, as characterized by the almost doubled ductility (~11.8±1.3% for the HET steel vs. ~6.2±0.6% for the HOM sample, Extended Data Fig. 11) under continuous electrochemical charging and by the greatly reduced (by a factor of two) area of brittle regions in the fracture surface of specimens loaded under H-plasma charging (Extended Data Fig. 12). These results suggest that the presence of a high density of Mn-rich buffer zones not only suppresses internal H migration and internal crack propagation, but also mitigates in-service H environmental embrittlement. The mechanism thus shows general effectiveness in enhancing a material's tolerance against intruding H, no matter how and when H exposure occurs and what the specific operating H embrittlement mechanism is (Extended Data Fig. 12).

The microstructure introduced here has been mainly architected to prove the proposed concept of chemical heterogeneity-induced crack arresting. More efficient crack blocking is likely to be achieved by further manipulating the number density and size of the $\gamma_{Mn-rich}$ regions via controlling thermomechanical processing parameters. The principle, based on the Griffith-type energy balance consideration, lies in that a larger Mn-rich buffer zone essentially requires a higher plastic energy for crack growth/opening, and a higher number density of $\gamma_{Mn\text{-}rich}$ reduces the initial size of H-induced microcracks which decreases the stress field at the crack tip.

To conclude, we turned chemical heterogeneity, normally undesired due to its detrimental effect on materials' conventional damage tolerance, into a mechanism that enhances the intrinsic resistance to H embrittlement. To avoid heterogeneities, alloys are usually subjected to costly high-temperature and long-time homogenization treatments that have a large negative environmental footprint. Our approach is counterintuitive: we designed and exploited chemical heterogeneity, rather than avoiding it, to arrest H-induced microcracks and suppress their propagation. The underlying thermodynamic principle, used here for developing microstructures with a specific degree of chemical heterogeneity, lies in the high kinetic mismatch between rapid phase transformation and sluggish solute diffusion, an effect generally applicable to alloyed steels [57,58]. Our approach can hence be transferred to a large variety of high-performance steels containing metastable austenite, targeting different application fields (e.g. various TRIP-aided advanced high-strength steels, metastable stainless steels and maraging-TRIP steels), using the same crack resistance-enhancing mechanisms introduced in this work. Further, our strategy of utilizing chemical heterogeneity is also expected to provide new insights into other advanced metal processing techniques such as powder metallurgy and additive manufacturing, where multiple options might exist to manipulate solute distribution/patterning [59-61]. In that context, the unique composite effect caused by solute heterogeneity, i.e. the combination of high crack resistance provided by local chemical fluctuations and high mechanical performance arising from other microstructure ingredients, can also be extended to other alloys in which a compositional dependence of H-resistance exists.

**Methods**

**Materials and processing**

The chemical composition of the material employed in this work is 0.2C-10.2Mn-2.8Al-1Si, in wt.% (0.8C-9.9Mn-5.5Al-1.9Si, in at.%). Its mass density was calculated to be ~7.5 g/cm$^3$ based on equations in Ref. [62], which is ~4% lighter than iron and ~6% lighter than common austenitic stainless steels. Casting was performed in a vacuum induction furnace. The steel ingots were then reheated to 1230 ˚C and hot rolled to ~3 mm above 750 ˚C, followed by air cooling to room temperature. The hot rolled plates consist of a predominately martensitic structure. The austenite's heterogeneous Mn distribution was realized through three processing steps after hot rolling: (1) A first intermediate annealing conducted at a low intercritical temperature (here 700˚C for 2 h) just above the Ac1 temperature (674 ˚C determined by dilatometry). This processing step allows the formation of some austenite ($\gamma_{Mn-rich}$) with high Mn enrichment (up to 16 wt.%) due to strong partitioning [63]. (2) A second-step cold rolling process with a thickness reduction of ~50%. This processing step has two aims. Firstly, it partially transforms $\gamma_{Mn-rich}$ to α'-martensite, accelerating the reversion of new austenite during the final annealing step [64]. Secondly, the partial austenite-to-martensite transformation fragments the $\gamma_{Mn-rich}$ zones, thereby increasing its number density and dispersion. The microstructure after this processing step is shown in Extended Data Fig. 1a. (3) The final intercritical annealing was carried out at 750 ˚C (50 ˚C higher than the preceding intermediate annealing temperature) for 5 min. This step enables the reversion of new austenite from martensite and ferrite. The new austenite formed at this stage contains a lower Mn content ($\gamma_{Mn-lean}$). Note that Steps 2 and 3 are essentially the same as in the established processing routes used for this type of alloy, and Step 1 can be readily implemented into current industrial practice (for example, by combining it with the coiling process).

Parameter selection for the final annealing treatment was guided by a diffusion-controlled transformation (DICTRA) simulation, using the TCFE7 and MOB2 databases. Since the initial stage of austenite reversion (controlled by rapid C diffusion and then Mn diffusion inside ferrite [65,66]) is much faster than Mn diffusion/homogenization inside austenite [65,66], sufficient reverted $\gamma_{Mn-lean}$ can form before a substantial Mn redistribution occurs between $\gamma_{Mn-lean}$ and $\gamma_{Mn-rich}$. The simulation results using the selected final annealing parameters (750 ˚C for 5 min) are shown in Extended Data Fig. 1b. It is demonstrated that compared with the pre-existed $\gamma_{Mn-rich}$, the newly reverted austenite ($\gamma_{Mn\text{-lean}}$) contains a lower Mn content controlled by the local equilibrium condition at the final annealing temperature. A heterogeneous Mn distribution inside the austenite phase is thus achieved.

The reference HOM sample was directly cold rolled to the same thickness reduction without the intermediate annealing step (as-cold rolled microstructure shown in Extended Data Fig. 2), followed by the final intercritical annealing at the same temperature (750 ˚C). The specimen treated by this "conventional" processing route shows a relatively homogenous Mn distribution inside austenite (Extended Data Fig. 3). Other microstructural characteristics including phase constituents, fraction and grain size are very similar for the HET and HOM samples (Extended Data Table 1).

**Hydrogen charging and mechanical testing**

The susceptibility to internal H embrittlement of the samples with and without austenite's chemical heterogeneity was evaluated by H pre-charging and slow strain rate tensile tests conducted at an initial strain rate of ~$8\times10^{-5}$ $s^{-1}$. Tensile experiments were carried out using a Kammrath & Weiss stage equipped with the digital image correlation (DIC) technique. Tensile specimens with a gage length of 4 mm and thickness of ~1.1 mm were used. Before tensile testing, electrochemical H charging was performed on tensile specimens in an aqueous solution containing 3 wt.% NaCl and 0.3 wt.% $NH_4SCN$ at room temperature. A platinum foil was used as the counter electrode. The total H concentration was controlled by varying the charging time (up to 48 h) and current density (5~15 $A/m^2$). The time interval between the end of H pre-charging and the start of tensile testing was less than 15 min. The H loss during this short time is negligible.

In order to evaluate the materials' susceptibility to H environmental embrittlement, we performed slow strain rate tensile tests under *in-situ* H charging to enable a continuous H ingress. Two types of *in-situ* charging methods with different H fugacities were used here. One is electrochemical H charging using the same electrolyte as that for H pre-charging and a current density of 5 $A/m^2$. The tests under this charging method were performed at an initial strain rate of $1\times10^{-5}$ $s^{-1}$ using a hydraulic testing machine (Zwick/Roell, Germany). Detailed experimental setup was described elsewhere [67]. Specimens with a gage length of 15 mm and width of 3 mm were used. The other charging method is H-plasma charging which was performed in a Quanta 650 FEG environmental scanning electron microscope (ESEM, ThermoFisher Inc., USA). Single-edge notched specimens (gage length 10 mm and width 4 mm) were loaded using a Kammrath & Weiss tensile module inside the ESEM, during which the plasma phase was ignited and continuously injected by an Evactron Model 25 Zephyr plasma source (XEI Scientific Inc., USA) using pure H gas as the process gas. More details about the H-plasma charging and its impact on the properties of ferritic steels can be found elsewhere [68,69]. Note that the partial pressure of the plasma phase in the current setup is about 40 Pa. This is a relatively low pressure in comparison with electrochemical cathodic charging (can reach tens of MPa of H fugacity [70]). Therefore, in this charging scenario, H ingress mainly occurs during the crack propagation stage which is promoted by the large amount of austenite-to-martensite transformation and high local stress/strain near the crack tip. Although the H influence on the macroscopic tensile properties is difficult to detect, this charging method provides a distinct contamination-free condition which is particularly useful for revealing detailed features of fracture surfaces [44,45].

It is important to mention that *in-situ* H charging during tensile testing occurs concurrently with the deformation-induced phase transformation from austenite to α'-martensite. Since the H diffusivity in α'-martensite is around 2~5 orders of magnitude higher than that in austenite [20,21,51,71], we expect a higher rate of H ingress at high strain levels, especially when most of the austenite has transformed to martensite and when high strain/stress is locally concentrated (e.g. near the tip of propagating cracks). In addition, the behavior of H trapping and migration upon H pre-charging or *in-situ* charging should be very different. For example, the initial microstructure of the investigated steel has a large area fraction of austenite-ferrite phase boundaries, which are strong H trapping sites (H binding energy ~50 kJ/mol [11]). However, such type of hetero-interface gradually disappears during deformation due to deformation-driven austenite-to-martensite transformation. This means that a high amount of H can be trapped at austenite-ferrite phase boundaries after H pre-charging [11], but this will not occur for the case of *in-situ* charging if H is mainly introduced at high inelastic deformation levels. The different behaviors of H ingress, trapping and migration could change the prevalent H embrittlement mechanisms, namely, from a dominant H-enhanced decohesion (HEDE) at hetero-interfaces [11] (Extended Data Figs. 7 and 10) for the case of H pre-charging, to a combined H-enhanced localized plasticity (HELP) and H-enhanced strain-induced vacancy (HESIV) mechanism (Extended Data Fig. 12) for the case of *in-situ* H-plasma charging. For either case, we demonstrate here, regardless of how and when H is introduced and what the specific operating H embrittlement mechanism is, that our approach of producing chemical heterogeneity always enhances materials' resistance to H embrittlement (Fig. 2 and Extended Data Figs. 11 and 12). In comparison to the role of $\gamma_{Mn-rich}$ in enhancing H-resistance in pre-charged samples (i.e. suppressing deformation-induced internal H migration as well as microcrack propagation), the architected $\gamma_{Mn\text{-}rich}$ zones have one additionally beneficial effect in resisting H environmental embrittlement, that is, they can also suppress the H ingress or penetration during deformation due to their trapping effect. This effect is reported to be more pronounced when most metastable austenite has transformed to martensite (i.e. the total austenite fraction achieves below 10 vol.% [72]).

**Thermal desorption spectroscopy and permeation testing**

Thermal desorption spectroscopy (TDS) experiments were performed using a Hiden TPD Workstation to measure the H concentration in pre-charged specimens. Specimens with a dimension of 10×15×1.1 mm$^3$ were used, which were subjected to the same H pre-charging condition as that for tensile specimens. The tests were started within 15 min after H charging. The total H concentration was determined by measuring the cumulative desorbed H from room temperature to 800 °C. The H diffusivity in the investigated samples was studied by permeation tests using a highly sensitive scanning Kelvin probe method [73]. Membranes with a thickness of 80~90 μm were prepared from the steel sheets to be probed, which were then coated with a ~100 nm Pd layer by physical vapor deposition (PVD, Leybold Univex 450) on the detection side. The H permeation through the specimen immediately enters into the Pd layer due to the lower chemical potential of H in Pd than in steel. The increase of the H concentration in the Pd layer then leads to a reduction in the surface potential which can be very sensitively measured by the Kelvin probe method [73]. The permeation curves obtained by this method can be used for calculating the effective diffusion coefficient of H, which is directly comparable with the traditional Devanathan-Stachurski method [73,74]. The effective diffusion coefficient, determined from the breakthrough time [72], is 6.0(±1.7)×10$^{-14}$ m$^2$/s for the non-deformed HOM specimen and 8.3(±1.2)×10$^{-14}$ m$^2$/s for the non-deformed HET specimen. The results reveal a similar H diffusivity in the two specimens, or even slightly faster H diffusivity in the HET sample due to its slightly larger grain size (lower interface density, i.e. 4533 mm/mm$^2$ for the HET sample vs. 5013 mm/mm$^2$ for the HOM sample measured from EBSD). Therefore, a similar penetration depth of H would be achieved for the two samples (or even slightly deeper for the HET sample) at a same H pre-charging condition.

**Microstructure characterization**

The scanning electron microscopy (SEM) observation was conducted using a Zeiss Sigma 500, a Zeiss-Merlin and a JEOL JSM-6500F field emission SEM (Zeiss Sigma 500 and Zeiss-Merlin for secondary electron (SE) imaging; Zeiss Sigma 500 and JEOL JSM-6500F for electron backscatter diffraction (EBSD)). EBSD data was analyzed using the TSL OIM software package. SEM-based Energy-Dispersive X-Ray Spectroscopy (EDX) was used to provide an overview of Mn distribution. It was conducted in a Hitachi SU-8230 cold-emission SEM, which was operated at an accelerating voltage of 10 kV using a Bruker Flat Quad 5060 F annular silicon drift detector (SDD). This high collection efficiency detector [75] was used to provide high signal-to-noise ratio x-ray images to improve the detection limit for Mn and other elements of the alloy. The weight fraction values in EDX mapping were further computed using the f-ratio method which was calibrated based on Monte Carlo modelling results [76,77]. The number density of $\gamma_{Mn-rich}$ per unit area was measured based on the SEM-EDX mapping, and it was then converted to the value per unit volume using the relation in Ref. [78] and a measured maximum size of $\gamma_{Mn-rich}$ (~0.5 μm), i.e. the value shown is the lower bound.

Scanning transmission electron microscopy (STEM) imaging and STEM-EDX were performed in a probe aberration-corrected TEM/STEM (FEI Titan Themis 60-300) with an acceleration voltage of 300 kV. For high angle annular dark-field (HAADF) imaging, a probe semi-convergence angle of 17 mrad and inner and outer semi-collection angles ranging from 73 to 200 mrad were used. High resolution imaging and selected area electron diffraction (SAED) were carried out in an image aberration-corrected TEM (FEI Titan Themis 80-300) operated at 300 kV. Thin foils for STEM and transmission Kikuchi diffraction (TKD) were prepared by a site-specific lift-out procedure [79] using a dual-beam focused ion beam (FIB) instrument (FEI Helios Nanolab 600i). TKD analysis was conducted using a FEI Scios FIB microscope at 20 kV.

**Statistical analysis of H-induced cracks**

After H pre-charging and tensile fracture, all the visible nano-/micro-cracks at regions adjacent to the fracture surface were probed by secondary electron imaging at an acceleration voltage of 5 kV. The samples were slightly etched by 1% nital to aid the identification of crack formation sites. In total around 150 and 250 cracks were probed for the HOM and HET samples pre-charged with ~6.5 wt ppm H, respectively. The formation sites (e.g. at ferrite/strain-induced α'-martensite interfaces or inside the α'-martensite) of these cracks were identified and their length was measured. The results are shown in Extended Data Fig. 10.

**Atom probe tomography**

Atom probe tomography (APT) was carried out using a LEAP™ 5000X HR instrument (Cameca Instruments Inc.) equipped with a reflection lens. Site-specific APT tips were prepared using a dual beam SEM/FIB instrument (FEI Helios Nanolab 600i) via an *in-situ* lift-out procedure [80]. Before APT measurements, TKD was performed on the polished APT tips inside the SEM (FEI Helios Nanolab 600i). APT experiments were operated in a laser pulse mode with the pulse rate of 200 kHz, pulse energy of 45 pJ, specimen temperature of 60 K and detection rate of 10 ions per 1000 pulses. The analysis of the APT data was performed by the IVAS 3.8.4 software.

**H segregation at the interface between ferrite and deformation-induced martensite**

An isocomposition surface with a threshold of 6 at.% Mn was first superimposed on the point cloud obtained from the dataset displayed in Fig. 4c, in order to delineate the interface between ferrite and deformation-induced martensite (transformed from $\gamma_{Mn-lean}$). A composition profile in the form of a proximity histogram was then calculated to estimate the composition near the interface [81]. The profile of Mn and H are plotted in Extended Data Fig. 7 in blue and green, respectively. There appears to be a slight segregation of Mn at the interface, but also a peak of H. The two horizontal lines correspond to the average composition in H in α and α', i.e. 0.82 at.% and 1 at.% respectively, and the peak composition, at the interface, is approx. 1.2 at.%.

Measuring the composition of H within APT datasets is extremely challenging and typically avoided, unless isotopic labelling is used [5,16]. Indeed, deuteration helps avoid overlap with the residual H from the vacuum chamber, which is typically found to affect data in regions of the specimen where the electrostatic field is low. However, recent studies have shown that specimen preparation by focused-ion beam leads to an uncontrolled incorporation of H within the specimen [5,82,83], which might be used as a qualitative assessment of the trapping/segregation behavior of H atoms. Interpreting variations in the measured H composition requires a careful assessment of the changes in the intensity of the electrostatic field used to trigger the field evaporation during the APT analysis [5]. Within the dataset (Fig. 4c in the manuscript), Fe appears in the 1+ and 2+ charge states, and the ratio of the charge states give qualitative information on the local changes in the electrostatic field, as studied in details by Kingham [84]. The log of the ratio of $Fe^{1+}$ to all charge states is plotted in dark orange along with the profile in Extended Data Fig. 7. Based on this data, we can see that the electrostatic field is slightly higher in α' than in α, and it drops slightly at the interface. Based on the recalculated Kingham curves in Ref. [85], the difference in the electric field between the two phases is around 2%. The higher amount of H in α', i.e. where the field is higher, can hence be safely interpreted as an indication that the H is effectively more highly concentrated in this phase [86], as can be expected due to the high dislocation density inside α'. Since the changes in the electrostatic field are very low, the peak of H at the interface can be attributed to an actual segregation of H at the α-α' phase boundary, no matter whether the H comes from electrochemical charging or from focused-ion beam. Such observation validates that H tends to segregate at martensite associated hetero-interfaces which promotes interface decohesion as shown in Fig. 4b and Extended Data Fig. 10a.

**Acknowledgments:** We thank N. Brodusch for the assistance in performing SEM-EDX experiments. S. Vakili, J.R. Mianroodi and P. Shanthraj are acknowledged for insightful discussions. F. Fazeli and C. Scott are acknowledged for their help in the preparation of materials. B.S. is grateful for the research fellowship provided by the Alexander von Humboldt Foundation; **Author contributions:** B.S. conceived the project and designed the experimental/modeling program. D.P. and D.R. supervised the study. B.S. was the lead experimental scientist of the study. W.L conducted TEM experiments. S.K.M. performed site-specific sample preparation for APT. S.K.M. and B.S. conducted APT and TKD experiments. B.S. and B.G. analyzed the APT data. C.W performed the permeation tests. D.W. performed tensile testing under *in-situ* H-plasma charging and associated fracture surface characterization. R.D., H.C. and B.S. performed the DICTRA simulation. B.S., B.G., and D.R. wrote the manuscript. All the authors discussed the results and commented on the manuscript. **Competing interests:** Authors declare no competing interests; **Data and materials availability:** All data is available in the main text or the extended data. Correspondence and requests for materials should be addressed to B.S.